\documentstyle[twoside,fleqn,espcrc2]{article}
\input boxedeps.tex
\SetRokickiEPSFSpecial  
\HideDisplacementBoxes

\newcommand{\D}{\overline{D}}
\newcommand{\da}{\dagger}  
\newcommand{\be}{\begin{equation}}
\newcommand{\eq}{\end{equation}}
   
\newcommand{\Tr}{{\rm \, Tr \!}}    

\newcommand{\AmS}{{\protect\the\textfont2
  A\kern-.1667em\lower.5ex\hbox{M}\kern-.125emS}}

\begin{document}

\title{Light-cone QCD on the lattice\thanks{Work 
presented by SD at 
Lattice '99, Pisa, Italy, and by BvdS at NuSS '99, Kyungju, Korea.}
\vskip-3cm\hfill\small DAMTP-99-103\vskip2.6cm
}
\author{S. Dalley\address{D.A.M.T.P., Silver Street, Cambridge CB3
9EW, U.K.},
Brett van de Sande\address{Institut fur Theoretische Physik III,
Universitat Erlangen, Germany}}

\begin{abstract}
Ideas and recent results for light-front
Hamiltonian quantisation of lattice gauge theories.
\end{abstract}
\maketitle
\section{Introduction}

A number of physicists have urged the development of
a workable light-front (LF) Hamiltonian formulation of QCD 
\cite{stanetc}. Such
a quantisation scheme would yield Lorentz-boost-invariant
wavefunctions having a natural constituent structure,
providing a power tool in the analysis of hadronic physics.
The most obvious  applications include inclusive and exclusive
hard scattering, and weak and electromagnetic decays, but
many new possibilities --- yet undreamt of ---
would be opened up with such a dynamical
framework.

So why has nobody done it? The same dynamical reasons that lead to
simplified wavefunctions, in particular the infamous triviality of the
vacuum state in the presence of high-energy cut-offs, also 
complicate the construction of renormalised LF Hamiltonians.
Non-perturbative effects normally associated with the vacuum
must appear explicitly in the Hamiltonian. This limits the scope 
of a traditional perturbative RG analysis (see, for example,
Refs.~\cite{ohio} for attempts to formulate weak-coupling
LF RG's). Recently, we have attempted to calculate renormalised
LF Hamiltonians in lattice gauge theory, using gauge and
Lorentz symmetries of low-energy observables to non-perturbatively 
fix coupling constant trajectories \cite{dv}. So far, this method
has been used to study glueballs and the heavy-quark potential
in the large-$N$ limit. This talk will review the
elements of light-front quantisation of lattice gauge
theories, and present some recent results for glueballs.

\section{Transverse Lattice Hamiltonians.}
\label{construct}

In $3+1$ spacetime dimensions we introduce a square lattice of spacing $a$
in the `transverse' directions ${\bf x}=\{x^1,x^2\}$ 
and a continuum in the $\{x^0,x^3\}$ directions.
In light-front (LF)
coordinates $x^{\pm} = (x^0 \pm x^3)/\sqrt{2}$, we treat $x^+$
as canonical time and place anti-periodic boundary conditions
on $x^- \sim x^- + {\cal L}$.\footnote{Continuum 
Lorentz indices are denoted thus $\alpha,\beta \in \{+,-\}$
and transverse indices thus $r,s\in \{1,2\}$. Repeated indices are
summed.} 
 Both $1/a$ and ${\cal L}$ are
high-energy cut-offs for the LF Hamiltonian
$P^- = (P^0 - P^3)/\sqrt{2}$ that evolves the system in
LF time $x^+$. This is because the LF free-field dispersion
relation for a particle of mass $\mu$ 
\be 
P^- = {\mu^2 + |{\bf P}|^2 \over 2 P^+}
\eq
has LF energy inversely proportional to light-front momentum
$P^{+} = (P^0 + P^3)/\sqrt{2}$ (conjugate to $x^-$). 
Since $P^{+} \geq 0$ and is conserved, the LF vacuum, 
specified by total $P^+ = 0$, can only contain $P^+ = 0$ modes.
The choice of anti-periodic null-plane boundary conditions removes
them, but their effects must be recovered in the
cut-off Hamiltonian somehow. We propose to do this by
constructing general LF 
Hamiltonians $P^-$ invariant
under gauge transformations and those Lorentz transformations
unviolated by the cut-offs, then tuning the remaining
couplings to recover the Lorentz symmetries violated
by the cut-offs. Lattice gauge formulations are ideal for performing
such a procedure non-perturbatively.

Unfortunately, in the
usual formulation of lattice gauge theories \cite{wilson2}, 
with degrees of freedom
in $SU(N)$, it is not straightforward to identify the independent
degrees of freedom, which is essential for canonical Hamiltonian quantisation.
The tricks of `equal-time' Hamiltonian lattice gauge theory in
temporal gauge \cite{suss}
do not carry over to light-front Hamiltonian lattice gauge theory
in any convenient way \cite{griffin}.
However, one does not have to choose lattice variables in $SU(N)$.
It was noted long ago by Bardeen and Pearson \cite{bard1} that lattice
variables $M_r$ in the space of all complex $N \times N$ matrices were
physically more appropriate for a coarse lattice. In this case, there
is no problem in identifying the independent degrees of freedom. The
penalty is that one is too far from the continuum to use 
weak-coupling perturbation theory. But in the case of light-front
Hamiltonians, this is of limited use anyway. 

The gauge field degrees of freedom below the cut-offs are represented by
hermitian gauge potentials $A_{\alpha}({\bf x})$ and 
complex link variables $M_r({\bf x})$. 
On the transverse lattice,
$A_{\alpha}({\bf x})$ is associated with a site ${\bf x}$,
while $M_r({\bf x})$ is associated with the link from ${\bf x}$ to 
${\bf x} + a \hat{\bf r}$. (Each `site' ${\bf x}$ is in fact a 
two-dimensional plane spanned by $\{x^+, x^-\}$.)
These variables transform under transverse lattice gauge
transformations $U \in SU(N)$ as 
\begin{eqnarray}
        A_{\alpha}({\bf x}) & \to & U({\bf x}) A_{\alpha}({\bf x}) 
        U^{\da}({\bf x}) + {\rm i} \left(\partial_{\alpha} U({\bf x})\right) 
        U^{\da}({\bf x})  \nonumber \\
        M_r({\bf x}) &  \to & U({\bf x}) M_r({\bf x})  
        U^{\da}({\bf x} + a\hat{\bf r})  \; .
\end{eqnarray}
Since it will be 
possible to eliminate $A_{\alpha}$ by partial gauge-fixing,
$M_r$ represents the physical transverse polarizations.

The simplest gauge covariant combinations are 
$M_r$, $F^{\alpha \beta}$, $\det M$, $\D^{\alpha} M$,
{\em et cetera}, where the  covariant derivative is
\begin{eqnarray}
        \D_{\alpha} M_r({\bf x})  
        & =  & \left(\partial_{\alpha} +{\rm i} A_{\alpha} ({\bf x})\right)
        M_r({\bf x})  
\nonumber \\ &&  
-  {\rm i} M_r({\bf x})   A_{\alpha}({{\bf x}+a \hat{\bf r}}) 
\ .
\label{covdiv}
\end{eqnarray}
To proceed, we must make some
assumptions about which finite sets of operators to
include in a real calculation. 
%
%
%
The following criteria were used to select operators in $P^-$ for pure
gauge theory:
\begin{enumerate}
\item Quadratic LF momentum operator $P^+$
\item Naive parity restoration as ${\cal L} \to \infty$.
\item Transverse (lattice) locality
\item Expansion in gauge-invariant powers of $M_r$
\end{enumerate}

We do not have space here to explain each of these criteria in detail,
but note that the last three can
all be straightforwardly checked in principle by systematically
relaxing the condition.
In all calculations we explicitly extrapolate to the 
${\cal L} \to \infty$ limit\footnote{We also used a Tamm-Dancoff
cut-off
and extrapolated this as well.} at fixed $a$, deriving $P^-$ from a 
Lagrangian including only dimension 2 operators
with respect to $\{x^+ , x^- \}$ coordinates. 
After fixing to LF gauge $A_{-} = 0$ and eliminating
the resultant constrained field $A_{+}$,
Fock space states will consist of 
free link-partons $a^{\da}(k^+)$ obtained by Fourier analysing $M_r$ 
in the $x^-$ coordinate
\begin{eqnarray}
M_r &  =  & 
        \frac{1}{\sqrt{4 \pi }} \int_{0}^{\infty} {dk^+ \over {\sqrt{ k^+}}}
        \left( a_{-r}(k^+)\, e^{ -{\rm i} k^+ x^-}  
\right. \nonumber \\ && \vspace{0.25in} \left. 
+   a^{\da}_r(k^+)\, e^{ {\rm i} k^+ x^-} \right) \ . \label{fock}
\end{eqnarray}
For sufficiently large
lattice spacing $a$, $M_r$ remains a massive degree of freedom 
and there is an energy
barrier for the addition of a link-parton to a 
Fock state. This effect is very pronounced in LF gauge
theories \cite{igor} and motivates condition 4 above.
By expanding $P^-$ to a given order of $M_r$ in this regime --- the
colour-dielectric expansion --- we 
cut off interactions between lower-energy few-parton states and 
higher-energy many-parton states. 
The advantage
of a cut-off on changes of parton number is that it organizes
states into a constituent hierarchy, consistent with energetics.
This kind of expansion was suggested in Ref.~\cite{bard2}, but only
recently have we been able to check that it works in practice. 

To leading order, the large-$N$ Lagrangian density satisfying our
conditions is
\begin{eqnarray}
L_{\bf x} & = & \D_{\alpha} M_r({\bf x}) (\D^{\alpha} M_r({\bf x}))^\da 
\nonumber \\ 
&& 
- {1 \over 2 G^2} \Tr \ \{ F^{\alpha \beta}F^{\alpha \beta} \} - V_{\bf x}
\end{eqnarray}
where 
\begin{eqnarray}
 V_{\bf x} & = & 
-{\beta \over Na^{2}} \Tr\left\{ M_{\rm plaquette} +  M_{\rm plaquette}^{\da} \right\}   
\nonumber \\
&& +    
 \mu^2  \Tr\left\{M_r M_r^{\da}\right\} 
\nonumber\\
&& +   {\lambda_1 \over a^{2} N}
\Tr\left\{ M_r M_r^{\da}
M_r M_r^{\da} \right\} 
\nonumber\\ && +
 {\lambda_2 \over a^{2} N}  
\Tr\left\{ M_r ({\bf x}) M_r({\bf x} + a \hat{r} ) \right. \nonumber \\
&& \left. M_r^{\da}({\bf x} + a \hat{r} ) M_r^{\da} ({\bf x})\right\} 
\nonumber \\
&& +    {\lambda_3 \over a^{2} N^2} 
\left( \Tr\left\{ M_r M_r^{\da} \right\} \right)^2
\nonumber\\ &&
+  {\lambda_4 \over a^{2} N}  
\sum_{\sigma=\pm 2, \sigma^\prime = \pm 1}
        \Tr\left\{ 
M_\sigma^{\da} M_\sigma M_{\sigma^\prime}^{\da} M_{\sigma^\prime} \right\} 
        \nonumber\\
&& +  {4 \lambda_5 \over a^{2} N^2} 
\Tr\left\{ M_1 M_1^{\da} \right\}\Tr\left\{ M_2 M_2^{\da} \right\} 
\ . \label{pot}
\end{eqnarray}
Collecting everything together, the canonical momenta in LF
gauge $A_{-} = 0$ become
\begin{eqnarray}
P^+ & = & 2 \int dx^- \sum_{{\bf x}} \Tr  
                        \left\{ \partial_- M_r({\bf x})  
                \partial_- M_r({\bf x})^{\da} \right\} \nonumber \\
P^-  & = & \int dx^- \sum_{{\bf x}}  
                V_{{\bf x}} - \Tr \  \left\{ A_{+}({\bf x})
        J^{+}({\bf x}) \right\}  
\nonumber \\ &&  
- {1 \over G^2}
 \Tr \ \{ \partial_{-}A_{+}\partial_{-}A_{+} \} \\
J^{+}({\bf x}) &=&  {\rm i} \left(
M_r ({\bf x}) \stackrel{\leftrightarrow}{\partial}_{-} 
M_r^{\da}({\bf x})  \right. \nonumber \\
&& \left. + M_r^{\da}({\bf x} - a\hat{\bf r}) 
\stackrel{\leftrightarrow}{\partial}_{-} M_r({\bf x} - a\hat{\bf r})   \right)
\end{eqnarray}

$A_+$ is a non-dynamical variable in this gauge, and eliminating
it introduces non-local interactions thus
\begin{eqnarray}
 P^-  &  = &  \int dx^- \sum_{{\bf x}}
  {G^2 \over 4} \Tr\left\{ {J^{+} \over \partial_{-}} 
{J^{+} \over \partial_{-}} \right\} 
\nonumber \\ && 
- {G^2 \over 4N} \Tr \ \left\{ {J^{+} \over \partial_{-}}  \right\}
\Tr \ \left\{ {J^{+} \over \partial_{-}}  \right\} + V_{{\bf x}}
\end{eqnarray}
where $J^+ / \partial_{-} \equiv  \partial_{-}^{-1}(J^+)$.
There is still a residual $x^-$-independent gauge invariance
generated by the charge $\int dx^- J^+$. As
originally shown in Refs.~\cite{bard1,bard2}, finite energy states 
$|\Psi\rangle$ are
subject to the gauge singlet condition $\int dx^- J^+ |\Psi\rangle = 0$.
In the large-$N$ limit, this means that Fock space at fixed $x^+$ is formed
by connected closed loops of link variables $M_r$ on the transverse
lattice (the $x^-$ coordinate of each $M_r$ is unrestricted).

\section{Glueballs}

The dynamical problem is now to diagonalize $P^-$, at fixed total
momentum $P^+$,  in the Fock basis (\ref{fock}).
To test low-energy eigenfunctions of $P^-$ (i.e. glueballs) 
for Lorentz covariance,
and so determine the couplings appearing in (\ref{pot}), we
also need states at non-zero ${\bf P}$ and methods to determine
the lattice spacing $a$ and string tension $\sigma$ \cite{burk}. 
Further details
can be found in Refs.~\cite{dv}.

We applied a $\chi^2$ test for a range of observables that measure
violations of Lorentz covariance, including anisotropy of glueball
dispersion, asymmetry of the heavy-quark potential, and splitting
of Lorentz multiplets in the spectrum. A distinctive one-parameter trajectory
${\cal T}_s$ appears in the space of couplings, largely universal
with respect to the precise details of the $\chi^2$ test,
along which 
$\chi^2$ is greatly reduced. We believe, as a result,
that this is
an approximation to a fully Lorentz-covariant scaling trajectory
${\cal T}$ that
exists in the infinite-dimensional space of all Hamiltonians and
posses a continuum limit $a \to 0$. The lattice spacing, whose
value is deduced as part of the analysis, remains quite large
($\sim 0.5$ fm)
on the piece of ${\cal T}_s$ that we can investigate. 
Thus, ${\cal T}_s$ may encounter barriers to the continuum and, 
for this reason, we do not try to extrapolate to $a=0$.  Instead, we look
for approximate scaling behaviour 
on coarse lattices
and estimate systematic errors empirically from violations of Lorentz
covariance.

\begin{figure}
\TrimRight{135pt}
$\frac{\cal M}{\sqrt{\sigma}}$\BoxedEPSF{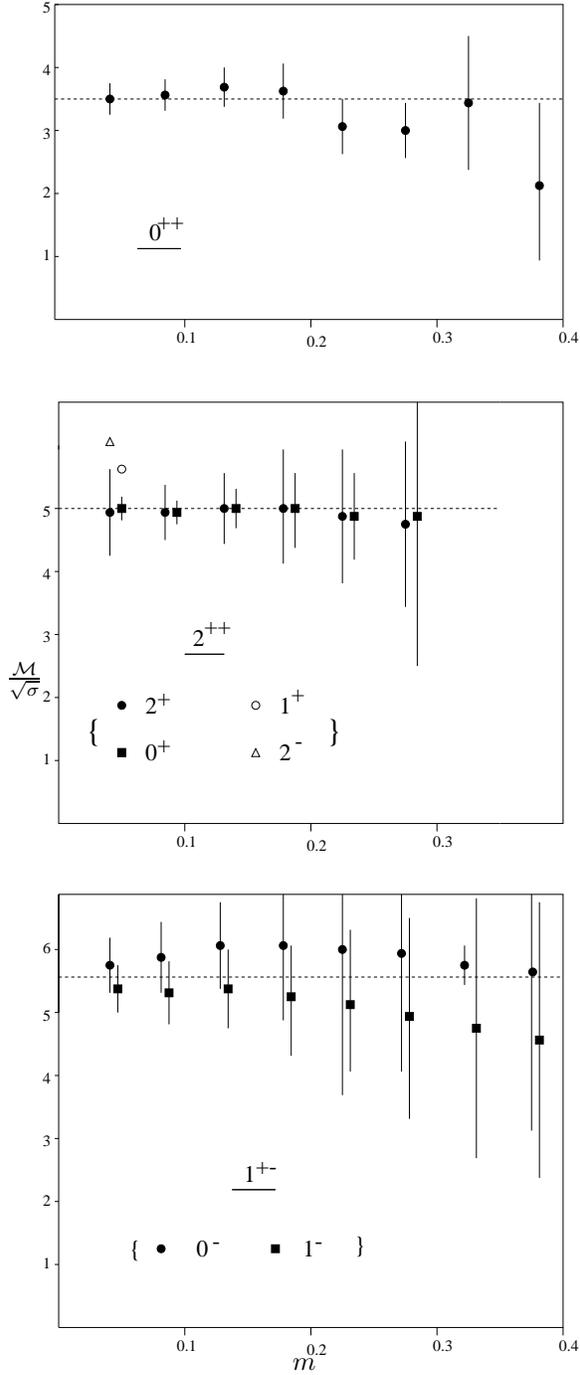 scaled 330}\\
\hspace*{1.5in}$m$
\caption{The variation of glueball masses with link-field mass $m$
along
${\cal T}_s$. The open-symbol data for the $2^{++}$
are still too 
inaccurate for error estimates. 
\label{scaling}}
\end{figure}

\begin{figure}
\centering
$\frac{\cal M}{\sqrt{\sigma}}$\BoxedEPSF{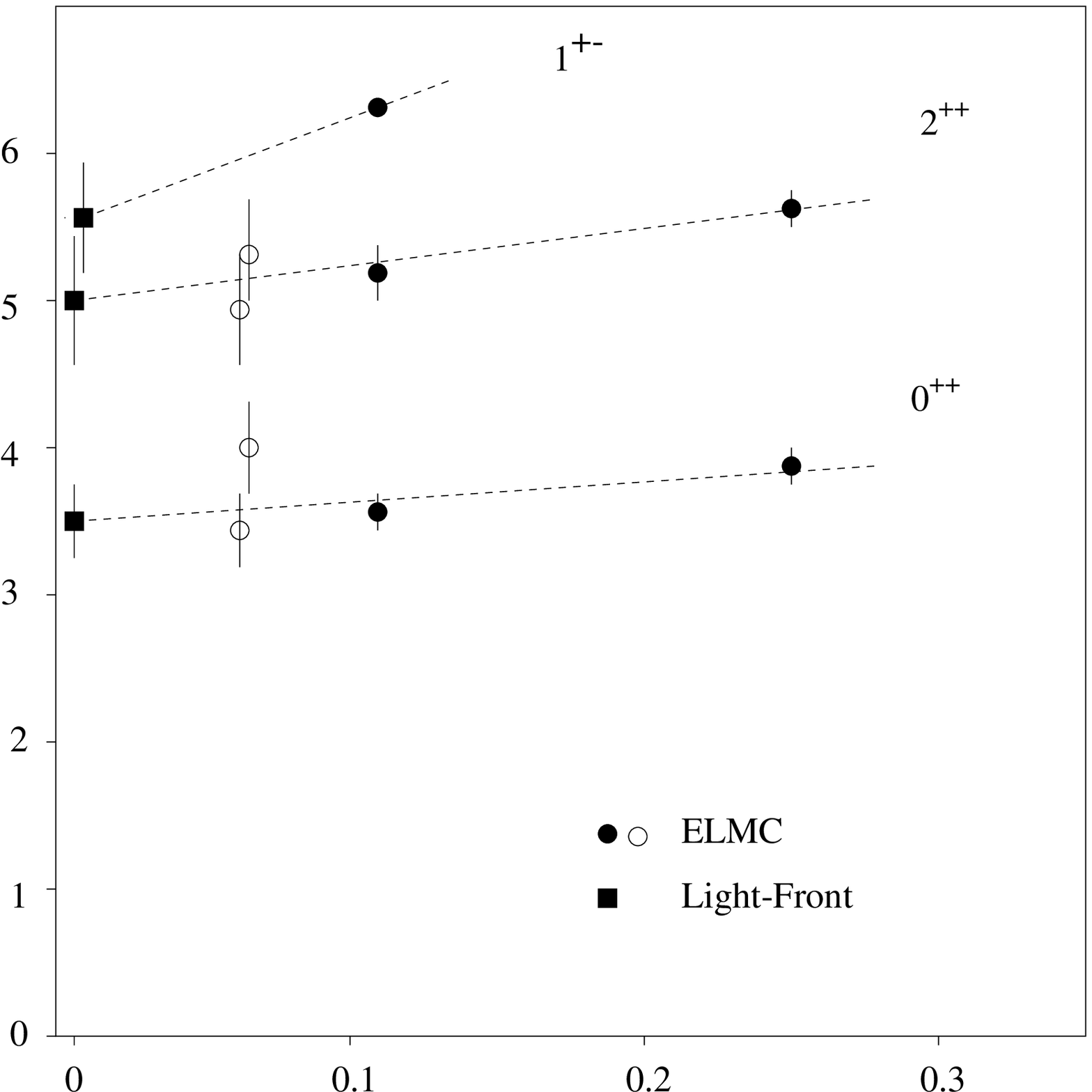 scaled 420}\\
\hspace{0.5in}$1/N^{2}$
\caption{The variation of glueball masses ${\cal M}$ with $N$ (pure glue). 
Euclidean Lattice Monte Carlo (ELMC) results are continuum estimates 
for $N=2,3$ \protect\cite{old,lattice,star} and 
fixed lattice spacing 
results for $N = 4$ \protect\cite{teper2}. 
The dotted lines are to guide the
eye, corresponding to leading linear dependence
on $1/N^2$.
\label{alln}}
\end{figure}

\begin{figure}
\centering
$G_d$\BoxedEPSF{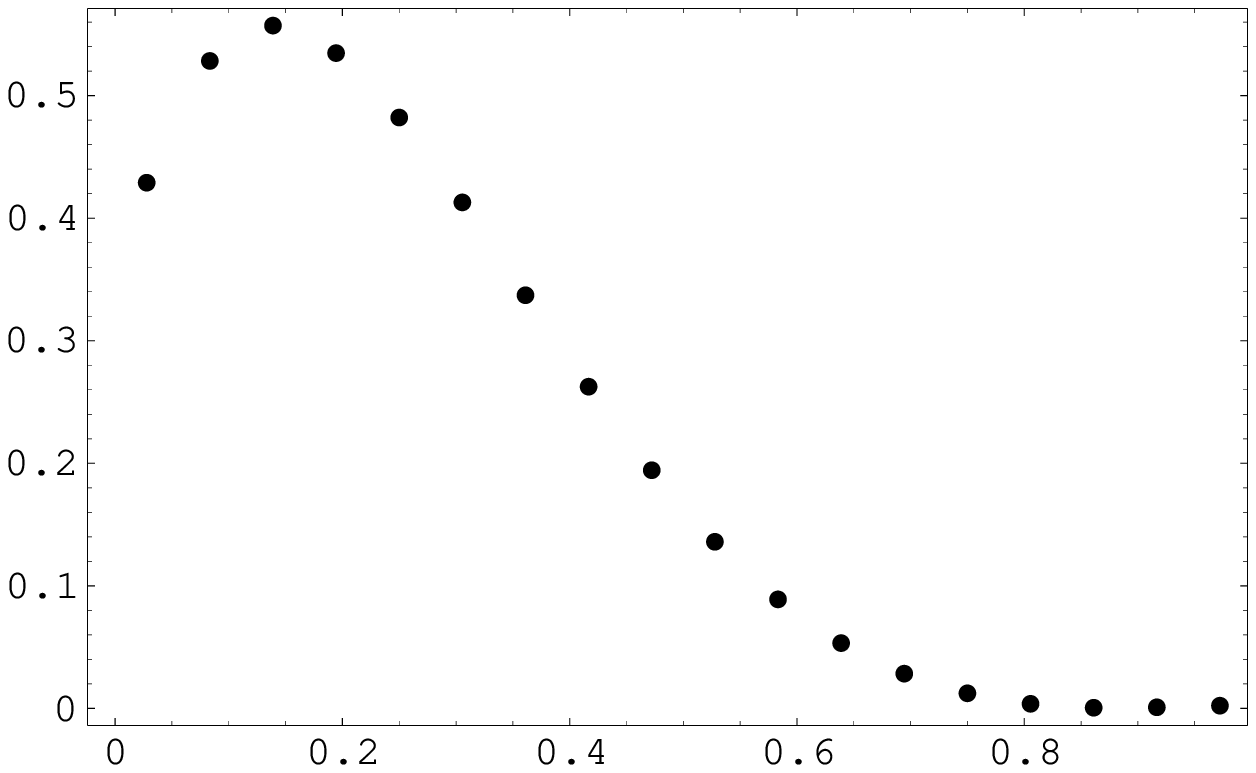 scaled 550}\\
\hspace{0.5in}$x$
\caption{The $0^{++}$ momentum distribution function at 
$a \sim 0.5$ fm.
\label{dist}}
\end{figure}

In fig.~\ref{scaling}
we plot the scaling behaviour of the lightest glueball masses
${\cal M}$ along
${\cal T}_s$.
The components are labelled by exact transverse lattice symmetries 
$|{\cal J}_3|^{{\cal P}_1}$ (where ${\cal P}_1: x^1 \to -x^1$) and
grouped into would-be Spin--Parity--Charge-Conjugation multiplets 
${\cal J}^{\cal PC}$. The various components of a Lorentz multiplet 
become rapidly more covariant,
measured by their isotropy and degeneracy, as the link-field mass
$\mu = m  G \sqrt{N}$ is reduced.

It is interesting to plot glueball masses in pure gauge theory versus
$1/N^2$, since this is supposed to be the relevant expansion
parameter about $N= \infty$ \cite{hoof}. In Figure~\ref{alln} the
$N= \infty$ data are taken from the best overall $\chi^2$ of our
calculation.
There is remarkably
little variation of glueball masses with the number of colours. 
It gives further support to the notion that
$N=3$ is close to $N=\infty$ in many situations. 
In particular, popular flux-tube and string models of
the soft gluonic structure of hadrons are typically more appropriate to the
large-$N$ limit of QCD (or are independent
of $N$). Figure~\ref{alln} indicates that these models should give 
worthwhile approximations to $N=3$ QCD .

From the explicit glueball eigenfunctions one can extract various
measurements of glueball structure.
An interesting quantity is the distribution of longitudinal
momentum $P^+$ among the link partons. In Fig.~\ref{dist} we plot the
quantity
\begin{eqnarray}
G_{d}(x)  & = & 
{1 \over 2\pi x P^+} \int dx^- {\rm e}^{-{\rm i} x P^+ x^-}
\nonumber \\ &&
\langle\Psi(P^+)|
        \Tr\left\{ \partial_{-} M_r \partial_{-} M^{\dagger}_r \right\}
|\Psi(P^+)\rangle \ , \nonumber
\end{eqnarray}
which measures the probability 
of finding a link-parton carrying
momentum fraction $x$ of the glueball momentum $P^+$.
It depends upon
the transverse normalisation scale through $a$, though our
approximation to ${\cal T}$ is too crude to reliably
see physical evolution of $G_{d}$ with scale.

$G_{d}$ is related to the gluon distribution; it becomes
the gluon distribution in the limit $a \to 0$. Moreover, since $M_r$ is
some collective gluon excitation and the momentum sum rule is
satisfied, one would naively expect the gluon distribution at a
general scale
$a$  to be softer than $G_{d}$. 
The $0^{++}$ glueball 
does not seem to look like simply a two gluon boundstate,
%
%
which would have a light-front distribution peaked at $x=0.5$. 
Once quarks are coupled to the
problem, distributions such as Fig.~\ref{dist}  
should have distinctive experimental signatures.

\section*{Acknowledgements}
SD is supported by PPARC Advanced Fellowship No.\ GR/LO3965. 
Computations were performed at the Ohio Supercomputer Center and
the HLRS Stuttgart.
Data files are available at Internet URL
{\tt http://www.geneva.edu/\~{}bvds}.

\end{document}